Title : Parallel DNA Synthesis : Two PCR product from one DNA template

Bhardwaj Vikash[1] and Sharma Kulbhushan[2]

[1]Email: vikashbhardwaj@ gmail.com

[1]Current address: Government College Sector 14 Gurgaon, Haryana, India

[2]Email: kulsviro@gmail.com

[2]Current address: MCS Group, Institute of Nuclear Medicine and Allied Sciences, Timarpur, Delhi, India

Abstract: Conventionally in a PCR reaction, Primers binds to DNA template in an antiparallel manner and template DNA is amplified as it is. Here we describe an approach in which First primer binds in a complementary parallel orientation leading to synthesis in parallel direction. Further reaction happened in usual way leading to synthesis of final DNA product having opposite polarity then the template used. We first time proved that synthesis of DNA can happen in a parallel orientation. We first time have shown that from a single DNA template, Two different related PCR products can be synthesized.

Our fundamental knowledge of DNA structure is based on the Watson – Crick model of DNA double helix which is able to explain key processes that are essential for life. Conformational polymorphism of DNA is now extending beyond the Watson - Crick double helix. In addition to the families of the A, B and Z forms (Double helices with antiparallel orientation of strands ) existence of DNA double helix with parallel strand (ps-DNA) was proven (1,2,3). For instance, Tchurikov *et al* in 1986 have founded that fragments of short dispersed and actively transcribed suffix non coding exon sequence element (from 48 to 88 bp) of *Drosophilla* are complementary in a parallel orientation to the fragment of the 5' non coding sequence in adult mRNA of the alcohol dehydrogenase gene (from +47 to + 88 bp). There are three insertion in each sequence as well as four non complementary pairs. Thus, the parallel DNA strands of the suffix (41 bp) and alcohol dehydrogenase (42 bp) was found to have 32 complementary nucleotides (4). Such mutual orientation of strands and non watson crick base pairing can arise at low pH, as a consequnce of Chemical modifications of nucleotide bases or sugar – phosphate back-bone or upon ligand binding (2). Mirror repeats able to form parallel stranded DNA with AT, GC pairs are abundant in genomes of various organisms (5). The base pairs schemes have been established with Raman and FTIR spectroscopy and then verified by NMR. In parallel stranded DNA, AT pairs with two hydrogen bonds were found to be of reversed Watson –Crick type. Bases of a GC pairs in parallel stranded DNA are somewhat shifted relative to one another and also form two Hydrogen bonds. Circular dichroism (CD) spectra of parallel stranded DNA with mixed AT/GC composition have no major distinction from those of related antiparallel DNA which has allowed one to suppose that there are no drastic differences in nearest neighbor base pairs interactions between parallel stranded and antiparallel stranded double helices (6). The speceficity of the interaction between the strand in Parallel DNA has also been studied and it is so high that a parallel probe as short as 40 bp is able to detect a specefic band in Southern blot hybridizations on whole genome DNA (7). Even gene specefic silencing using parallel complementry RNA has been reported in *Pseudomonas aeruginosa* and *Escherichia coli* (8,9). Veitia and ottolenghi has compared (blast) a paralleled sequence (complementary read in a parallel direction) of a typical Alu element (Genbank AI: HSU14567) with the non-redundant division of Genbank.

Several significantcant hits have been reported in this search (i.e. AF176315.1, Z96292, Z96488, L15253) some of them involving subtelomeric sequences. The surprisingly high scores and low P values (ranging between $10e^{-32}$ and $10e^{-06}$) obtained for the sequence alignments suggested the existence of a close link between these sequences far beyond a mere random convergence. In principle, there are no thermodynamic constraints preventing Parallel nucleic acid synthesis from taking place. The dNTPs used for a normal polymerization reaction can also serve a parallel reaction provided the enzyme is able to catalyse the nucleophilic interaction between 3'OH and a 5'PPP from nucleotides arranged in a parallle way with respect to template. However, several attempts to amplify L15253 by PCR using different pairs of primers were unsuccessful. This negative result does not prove at all that parallel sequences do not exist but  author has suggested caution when handling results derived purely from in silico database searches (10).

The thermal denaturation analysis of parallel DNA has shown that it melts at lower temperature than the corresponding antiparallel structure (11). This finding gives us a clue that taking Double stranded antiparallel  DNA as a template for PCR will not be possible as during anealing steps antiparallel DNA will anneal to itself without binding to parallel stranded complementary  primers. So we proposed the hypothesis that this reaction can be possible if we start reaction using single stranded DNA as a template. A 120 bp single stranded DNA sequence was custom synthesized at Sigma aldrich.  PCR was performed using different combination of  complementary Primers in different orientation ( Table 1). As expected in one of reaction we found that 120 bp DNA template was amplified in which first primers binds in a complementary parallel orientation leading to synthesis in parallel direction. In second step, after denaturation, second primer annealed  to new DNA synthesized in an antiparallel way. The further reaction has procedded as usual PCR reaction (Fig.1). Similarly as a control reaction, single stranded 120 bp DNA was amplified by conventional PCR in which both the primers binds in antiparallel complementary way (Fig. 2). DNA sequencing result has confirmed that as expected 120 bp DNA template has been amplified in two different PCR products and one of the PCR product reads in a parallel direction to template DNA.

We first time have proved that DNA synthesis can happen in a parallel orientation. Our future target is to prove that two different protein can be synthesized from one DNA template. It will be interesting to know whether these protein will be interacting to each others or not. This invention will be usefull in designing new DNA sequencing reaction, 5'RACE and 3'RACE. We hope that it will open new biotechnology/ bioinformatics analysis of various genomes which will be based on parallel stranded DNA. We are at just beginning and hoping to find more.

**Acknowledgement**: I am thankful to Harpreet Singh for providing me custom synthesized template DNA and Primers used in this study.

11. German M.W., Kalisch W.B. and Van de Sande H.J. Relative stability of Parallel and antiparallel stranded duplex DNA. *Biochemistry. 27, 8302-8306 (1988).*

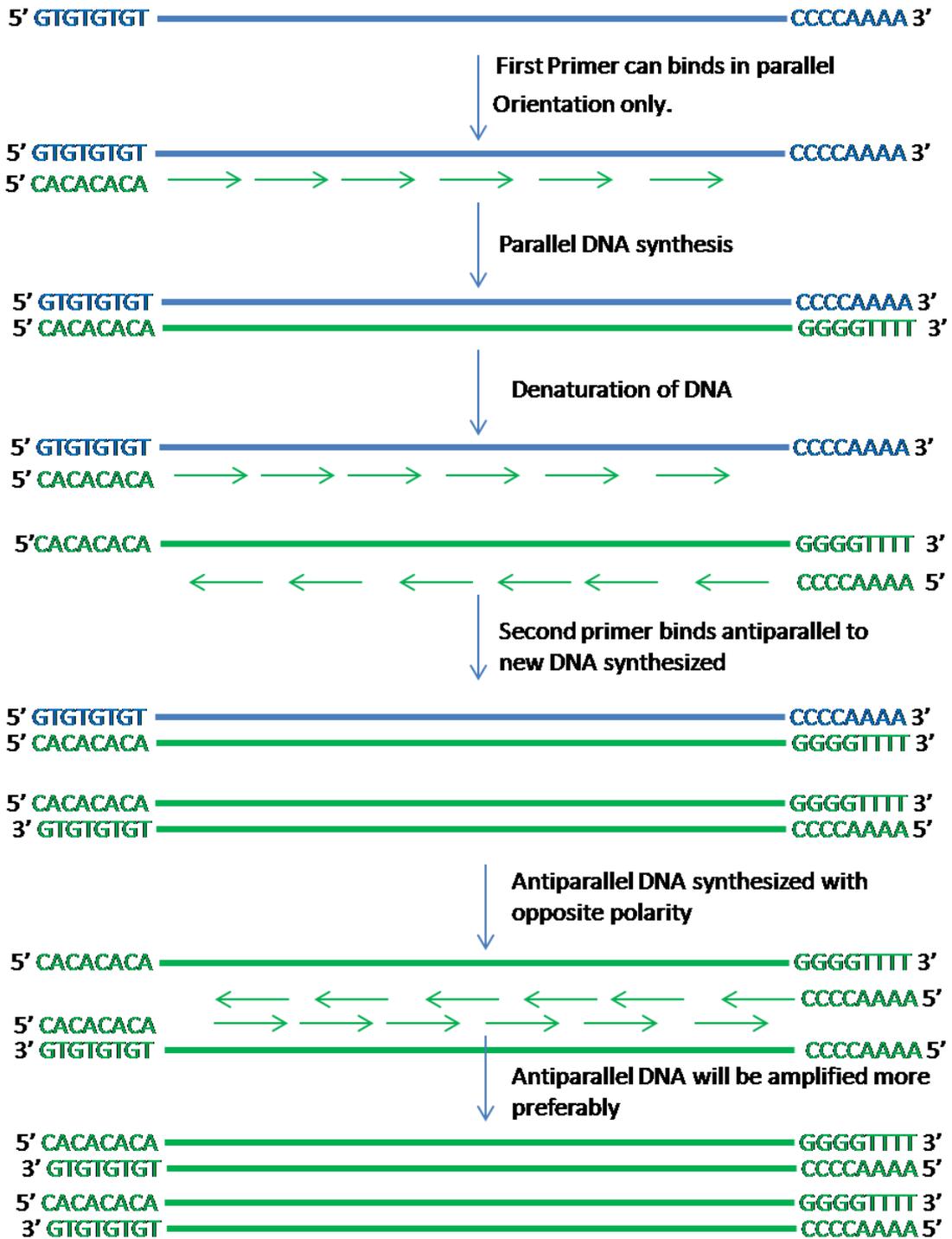

Figure 1: Flowchart showing scheme of Parallel DNA PCR. First primer binds in a complementary parallel manner to template DNA. The DNA syntesized runs in parallel orientation. After one complete PCR cycle , after denaturation during anealing step second primer binds to newly synthesized DNA in an antiparallel manner. Same set of primers now in a conventional way binds DNA in antiparallel complementary way leading to synthesis of DNA which will have opposite polarity then the template DNA.

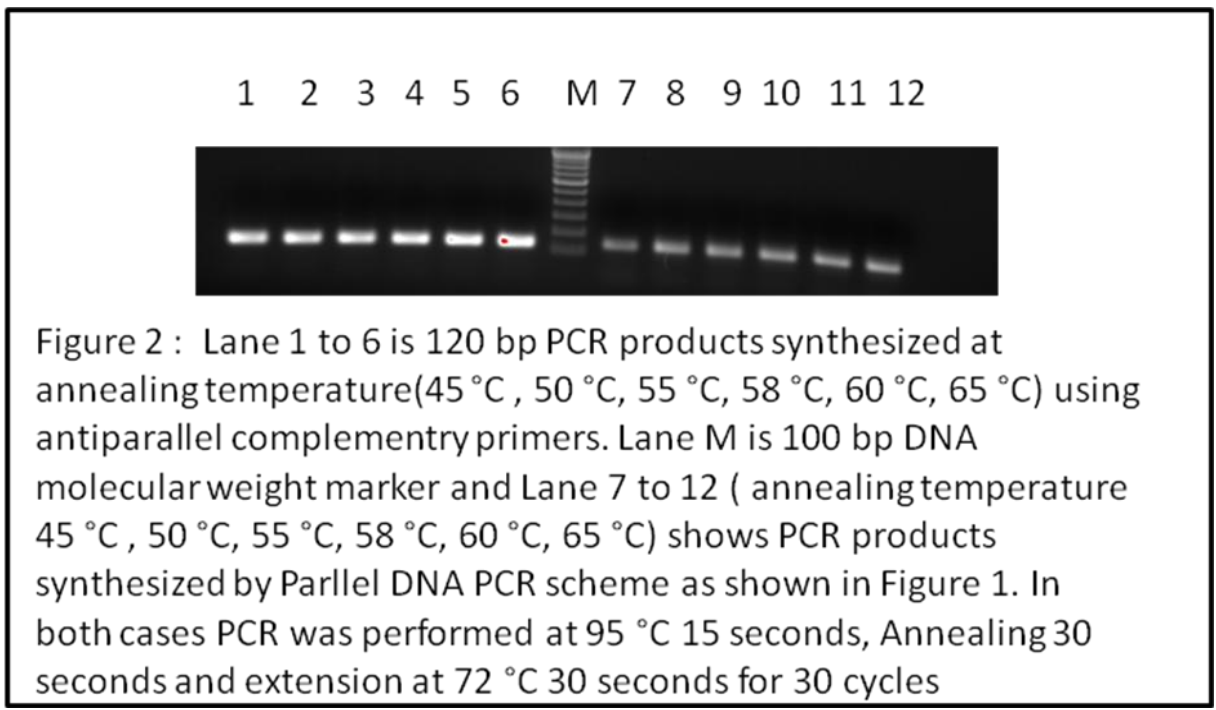

Figure 2 : Lane 1 to 6 is 120 bp PCR products synthesized at annealing temperature(45 °C , 50 °C, 55 °C, 58 °C, 60 °C, 65 °C) using antiparallel complementry primers. Lane M is 100 bp DNA molecular weight marker and Lane 7 to 12 ( annealing temperature 45 °C , 50 °C, 55 °C, 58 °C, 60 °C, 65 °C) shows PCR products synthesized by Parllel DNA PCR scheme as shown in Figure 1. In both cases PCR was performed at 95 °C 15 seconds, Annealing 30 seconds and extension at 72 °C 30 seconds for 30 cycles

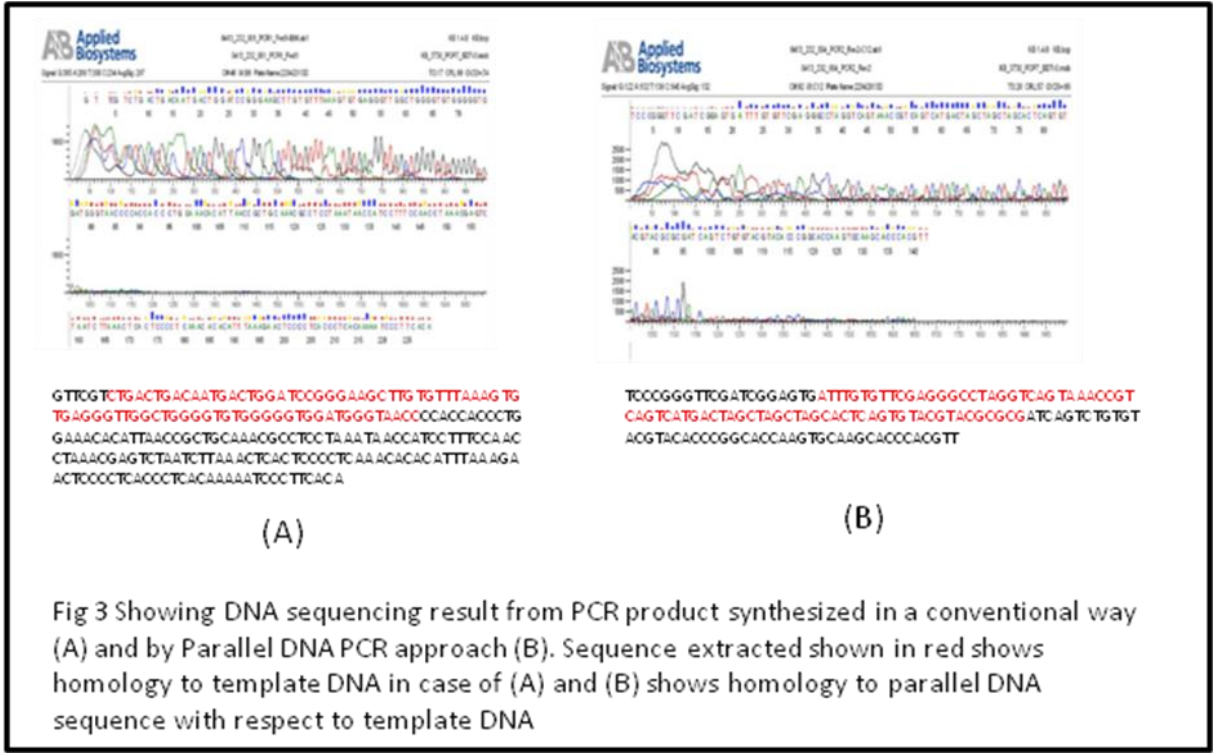

Fig 3 Showing DNA sequencing result from PCR product synthesized in a conventional way (A) and by Parallel DNA PCR approach (B). Sequence extracted shown in red shows homology to template DNA in case of (A) and (B) shows homology to parallel DNA sequence with respect to template DNA

| Template DNA | 5' GCG CGC ATG CAT GTG ACT GAC GAT CGA TCG ATC AGT ACT GAC TGA CAA ATG ACT GGA TCC GGG AAG CTT GTG TTT AAA GTG TGA GGG TTG GCT GGG GTG TGG GG G TGG ATG GGT AGC CGC 3' |
|---|---|
| PCR Primers<br><br>(Both Primers binds template DNA in antiparallel Manner ) | FWD Primer 1<br>5' GCG CGC ATG CAT GTG ACT GAC 3'<br><br>Rev Primer 2<br>5'GCG GCT ACC CAT CCA CCC CCA C 3' |
| Parallel DNA PCR Primers<br><br>(Primers binds to template DNA as scheme shown in Figure 1) | Fwd Parl 1<br>5'CGC GCG TAC GTA CAC TGA CTG C 3'<br>Rev Parl 2<br>5' CGC CGA TGG GTA GGT GG GGG 3' |

Table 1 : Showing custom synthesized DNA sequence used as a template DNA and Pair of primers.